# Ghost optical coherence tomography

Caroline Amiot[1,3*], Piotr Ryczkowski[1], Ari T. Friberg[2], John M. Dudley[3], and Goëry Genty[1]

[1]Laboratory of Photonics, Tampere University of Technology, Tampere, Finland

[2]Department of Physics and Mathematics, University of Eastern Finland, Joensuu, Finland

[3]Institut FEMTO-ST, UMR 6174 CNRS-Université Bourgogne-Franche-Comté, Besançon, France

*corresponding author

**Abstract:** We demonstrate experimentally ghost optical coherence tomography using a broadband incoherent supercontinuum light source with shot-to-shot random spectral fluctuations. The technique is based on ghost imaging in the spectral domain where the object is the spectral interference pattern generated from an optical coherence tomography interferometer in which a physical sample is placed. The image of the sample is obtained from the Fourier transform of the correlation between the spectrally-resolved intensity fluctuations of the supercontinuum and the integrated signal measured at the output of the interferometer. The results are in excellent agreement with measurements obtained from a conventional optical coherence tomography system.

Optical coherence tomography (OCT) is an interferometric imaging technique commonly used to obtain images of materials with high in-depth resolution [1-3] and it is commonly employed for industrial characterization of materials and components [4]. Due to its non-invasive character and the absence of ionizing radiation or toxic contrast agents in its implementation, OCT has also found multiple medical applications and it is now widely used in retinal, skin, and blood vessel examination [5-6]. A particular variant of OCT is performed in the spectral (wavelength) domain, based on measuring the phase relation between different wavelength components of a broadband light source in order to determine the distance to the sample under test [7]. A major benefit of spectral-domain OCT is that it does not require scanning along the sample direction allowing for significantly faster acquisition speeds compared to its time-domain counterpart. Spectral-domain OCT has been demonstrated using various types of light sources including broadband stationary and pulsed sources or swept-wavelength sources [8].

In parallel to the rapid development in OCT technologies, there has also been much recent interest in the unconventional imaging technique known as ghost imaging. Ghost imaging is based on the principle of image creation from the correlation between a known structured pattern that illuminates an object and the total integrated intensity transmitted (or reflected) by the object [9-10]. The defining feature of ghost imaging is that neither of the beams alone actually carries enough information to reconstruct the image. Rather, it is only by correlating the two measurements of the structured illuminating source and the integrated intensity from the object that an image can be generated. In this regard, the fact that the light actually detected from the object is an integrated intensity has also led to ghost imaging to be referred to as single-pixel imaging. A significant advantage of ghost imaging when compared to conventional imaging is that it is insensitive to distortions of the wavefront occurring after the object as only the total light intensity is measured [11], making it ideal for measurements in turbid media or in the presence of other noise. Ghost imaging can be performed using light sources with random spatial intensity patterns [12] or patterns which are controlled using e.g. programmable digital micro-mirrors [13-16]. Ghost imaging has also been extended to the temporal domain [17-18] and very recently in the frequency domain for real-time broadband greenhouse-gas spectroscopic measurements [19].

In this work, we combine the technique of OCT with the concept of spectral-domain ghost imaging to introduce a new methodology of ghost optical coherence tomography. This method generates a "ghost" spectral interferogram from the correlation between the spectrally-resolved intensity fluctuations of the light source and the integrated signal measured at the output of an OCT interferometer where a physical sample is placed. As in conventional OCT, the image of the sample is then retrieved from the Fourier transform of the interferogram. As a proof-of-principle demonstration, we (i) characterize the relative displacement of a perfectly reflecting mirror and (ii) perform measurement of the thickness of a microscope cover glass. The results are in excellent agreement with those obtained from a conventional OCT setup. A significant advantage of the ghost OCT scheme is that it does not require any particularly sensitive detector or spectrometer at the interferometer output. This could be extremely useful in situations where the object to be measured is highly absorbing or diffusing, for samples with low damage threshold or and for imaging in spectral regions where sensitive detectors are not available.

We begin by illustrating the concept of ghost spectral-domain OCT. To this end, Figure 1 compares the schematics of a conventional spectral-domain OCT system (Fig. 1a) and that of the ghost spectral-domain OCT approach (Fig 1b). In a conventional OCT system, the beam from a broadband light source is equally divided between the two arms of an equal path Michelson interferometer. The image of the object inserted in one arm is generated by measuring with a high-resolution spectrometer the spectral interference pattern resulting from the superposition of the beams reflected from the reference mirror and object.

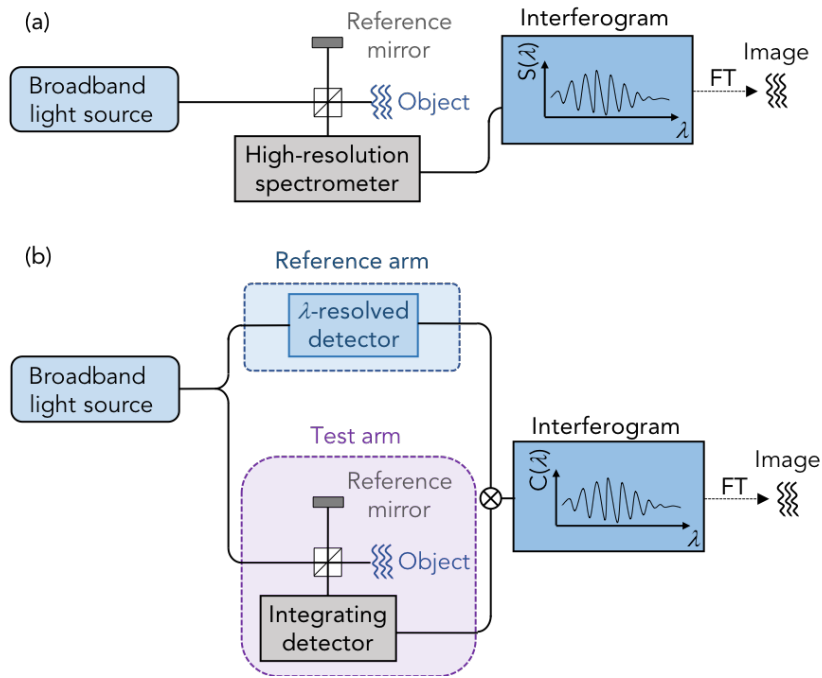

**Figure 1.** Schematic illustration of (a) a conventional spectral-domain OCT system and (b) the ghost spectral-domain OCT scheme. FT: Fourier transform.

The axial resolution of the system is inversely proportional the spectral bandwidth $\Delta\lambda$ of the source as $\delta z = 0.44 \lambda_0^2/\Delta\lambda$ (for a Gaussian spectral envelope), where $\lambda_0$ is the light source center wavelength. The imaging depth on the other hand is set by the spectrometer resolution. In spectral-domain ghost OCT, the beam from a light source with random spectral intensity fluctuations is divided between a reference arm where the fluctuations are measured in real time and a test arm consisting of a Michelson interferometer where the object to be measured is placed. The high-resolution spectrometer at the output of the interferometer is replaced by a slow integrating detector with no spectral resolution. The 'ghost' spectral interference pattern produced by the presence of object in the interferometer is then given by the normalised correlation function C(λ) between the reference and test arm signals defined by:

$$C(\lambda) = \frac{\langle \Delta I_{ref}(\lambda) \cdot \Delta I_{test} \rangle_N}{\sqrt{\langle \Delta I_{ref}(\lambda)^2 \rangle \langle \Delta I_{test}^2 \rangle}}. \quad (1)$$

Specifically, the normalized wavelength-dependent correlation function C(λ) represents correlation between multiple measurements of the spectral intensity fluctuations in the reference arm $I_{ref}(\lambda)$ and the total (or integrated) wavelength-independent intensity $I_{test}$ in the test arm at the output of the interferometer. $\langle \ \rangle_N$ denotes ensemble average over distinct $N$ realizations, and $\Delta I = I - \langle I \rangle_N$. The axial resolution $\delta z$ of the ghost OCT scheme is identical to that of the conventional OCT and the imaging depth is now determined by the resolution with which the spectral intensity fluctuations are measured.

Figure 2 shows our experimental setup. The light source is a spectrally incoherent SC extending from ~1300 to over 1700 nm and generated by launching 1 kW, 700 ps pulses at 1547 nm with 100 kHz repetition rate (Keopsys-PEFL-K09) into a 6-m long dispersion-shifted fiber (DSF) with zero-dispersion wavelength at 1510 nm (Corning Inc LEAF). The SC generating dynamics arise from noise-seeded modulation instability and soliton dynamics [20] resulting in large and random shot-to-shot spectral fluctuations across the entire SC spectrum [21]. These spectral fluctuations produce the (random) structured patterns that are used to probe the spectral OCT interferogram. Light from the SC source is divided between the reference and test arms with a 99/1 fiber coupler. In the reference arm, the spectral fluctuations are measured in real time using the dispersive Fourier transform technique [22-23]. Specifically, the single-shot spectra are converted into the time domain by a 150 km dispersion compensating fiber (DCF, FS.COM customized 150 km) with total dispersion of 3000 ps×nm$^{-1}$ and measured with 0.2 nm resolution using a 1.2 GHz InGaAs photodetector (Thorlabs DET01CFC/M) and 20 GHz real-time oscilloscope (Tektronics DSA72004) with 6.5 GS/s sampling rate. In order to avoid spectral distortion in the DCF due to attenuation and third-order dispersion, we restrict the SC bandwidth to the 1610–1670 nm range with a bandpass filter (Spectrogon NB-1650-050) corresponding to an axial resolution of ~20 µm. In the test arm, light is directed to a standard interferometer whose output intensity is measured with a slow photodetector (15 MHz bandwidth, Thorlabs PDA10D2). We emphasize that this simple detector cannot resolve the spectral interference pattern arising from an optical path difference between the two arms of the interferometer. The single-shot SC spectra from the reference arm and the corresponding spectrally integrated signal from the interferometer are recorded simultaneously by the oscilloscope. The correlation between these two computed over multiple realizations then yields the spectral interferogram whose Fourier transform gives the optical path difference between the interferometer arms.

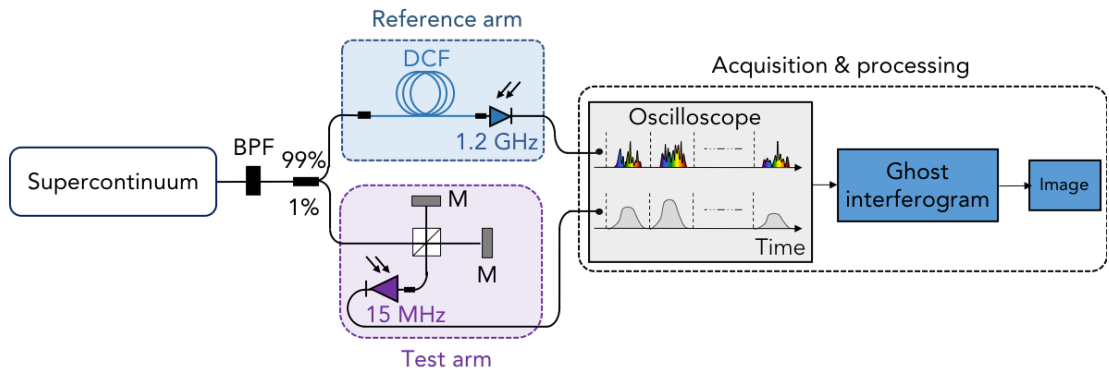

**Figure 2.** Experimental setup. BPF: band-pass filter, DCF: dispersion compensating fiber. M: mirror.

We first demonstrate the operating principle of ghost OCT by displacing the position of one of the interferometer mirrors from the zero-path difference position. The OCT interferogram is generated from the correlation function between the spectrally-integrated intensity measured at the interferometer output and the intensity of the single-shot SC spectra. The optical path difference between the two interferometer arms is then simply obtained from the Fourier transform of the interferogram. The results are plotted in Fig. 3(a) for increasing optical path difference between the mirrors. For comparison, we repeated the same measurements using the conventional OCT scheme by replacing the slow detector at the output of the OCT interferometer with a high-resolution optical spectral analyzer (OSA). The interferogram in this case is measured directly by the OSA and the reference arm measurements are not needed. The results are shown in Fig. 3(b) and we can see the excellent correspondence between the ghost and conventional OCT schemes across the full range of measured optical path differences.

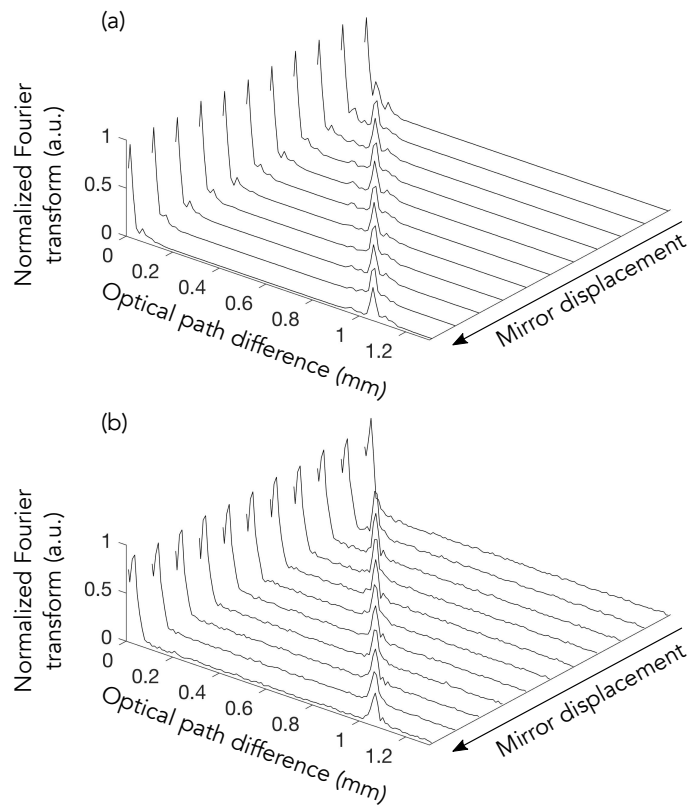

**Figure 3.** Optical path difference between the two arms of the Michelson interferometer measured by (a) the ghost OCT configuration and (b) conventional OCT. The arrow indicates the direction of the mirror displacement increasing the optical path difference.

In ghost imaging, the signal-to-noise (SNR) increases with the number of realizations used to compute the correlation, and we illustrate this in Fig. 4 where we show how the Fourier transform of the interferogram evolves as a function of the number of distinct SC pulses for a specific optical path difference (c.a. 0.55 mm in this case) in the interferometer. We can see that even with as few as 200 realizations corresponding to a total recording time of less than 2 ms, and although the SNR is not especially high, the interference fringes are already visible and the optical path difference can be retrieved. For a larger number of realizations, the SNR increases with the square of the number of realizations.

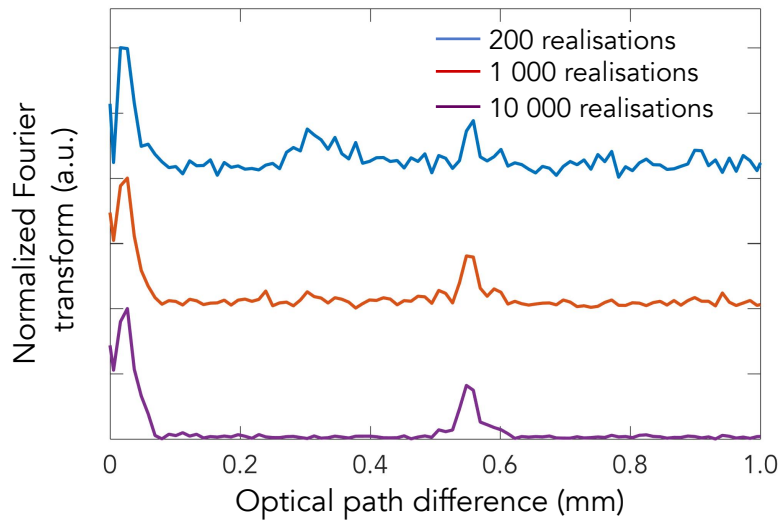

**Figure 4.** Normalized Fourier transform of the ghost interferogram for an optical path difference of 0.55 mm between the two arms of the interferometer and for an increasing number of realizations as indicated.

We next performed a second series of experiments using a dual-interface sample. For this purpose, we replaced one of the mirrors in the interferometer by a 210 μm thick (optical thickness) microscope cover slip consisting of two air-glass interfaces and the optical path difference between the two arms was set to be c.a. 1mm. The Fourier transform of the resulting interferogram is shown in Fig. 5 both for the conventional OCT system (Fig. 5a) and the ghost OCT setup (Fig. 5b). The results are again in excellent agreement and the positions of the two air-glass interfaces can be clearly identified. The distance between the two interfaces is measured to be 220 μm (optical thickness) close to the nominal value of 210 μm provided by the manufacturer.

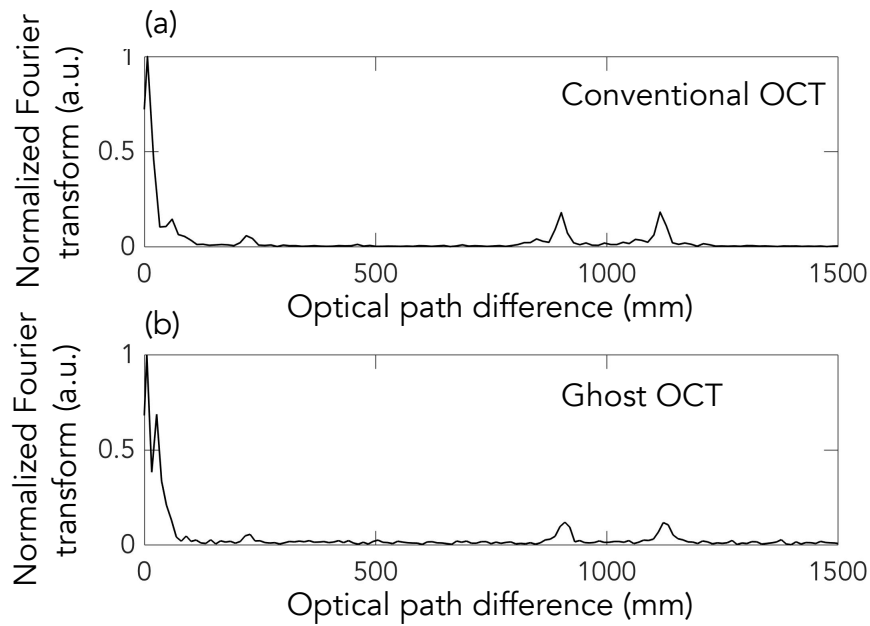

**Figure 5.** Optical path difference measured by (a) conventional OCT and (b) the ghost OCT configuration when one of the mirrors is replaced by a 210 μm microscope cover slip (optical thickness).

In conclusion, we have experimentally demonstrated proof-of-concept ghost OCT in the spectral domain using a broadband, spectrally incoherent, supercontinuum source.

As in a conventional OCT setup, the resolution is determined by spectral bandwidth of the source. The imaging depth on the other hand is given by the spectral resolution with which the spectral fluctuations of the light source can be measured in real time (and therefore the total dispersion if fiber dispersive Fourier transform is used). No particularly sensitive detector or spectrometer is needed at the interferometer output, which could be a significant advantage when the object is highly absorbing or diffusing, when the sample under has a low damage threshold and does not tolerate high intensity, or for imaging in spectral regions where sensitive detectors are not available. Finally, we note that the method can be implemented both with classical light sources and entangled photon sources and that a computational version may be realized by using e.g. controllable frequency combs which would eliminate for the need of single shot spectral measurements to perform the correlation.

**Acknowledgments:** C.A. acknowledges the support from TUT and SPIM graduate schools. J.M.D. acknowledges support from the French Investissements d'Avenir program, project ISITE-BFC (contract ANR-15-IDEX-0003). G.G. acknowledges the support from the Academy of Finland (grant 298463).